\documentclass[aps,prl,10pt, twocolumn, superscriptaddress, nobibnotes]{revtex4-2}

\usepackage{graphicx}
\usepackage{amssymb}
\usepackage{amsmath}
\usepackage{color}
\usepackage{graphicx}
\usepackage{dcolumn}
\usepackage{epstopdf}
\usepackage{bm}
\usepackage{tikz}
\usepackage{siunitx}
\usepackage{soul}
\usepackage[colorlinks=true,citecolor=blue]{hyperref}

\newcommand{\orcid}[1]{\href{https://orcid.org/#1}{\includegraphics[width=8pt]{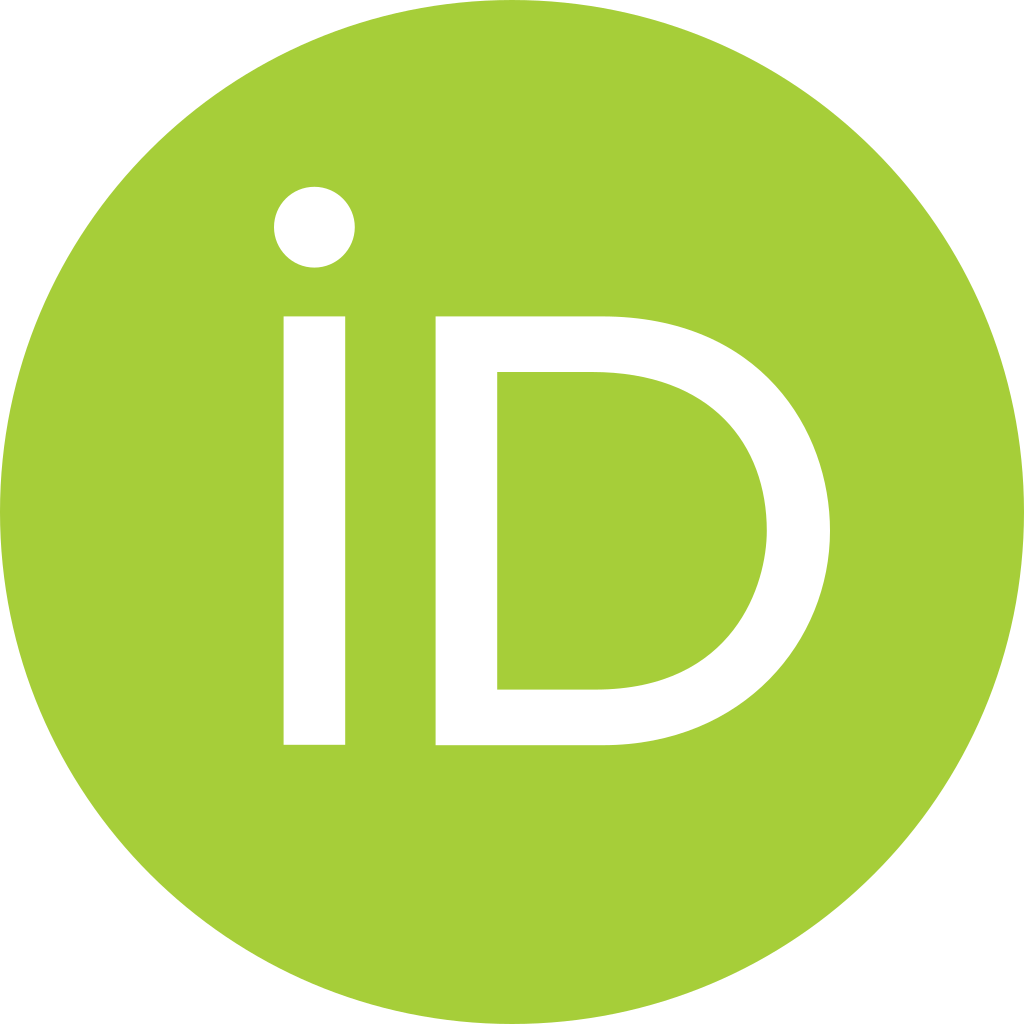}}}

\begin{document}

\title{Protected Charge Readout Scheme Utilizing Nonlinear Resonators}

\author{Andrea Cicovic\orcid{0009-0009-7402-2054}}
\email{andrea.cicovic@fysik.lu.se}
\affiliation{NanoLund and Solid State Physics, Lund University, Box 118, 22100 Lund, Sweden}
\author{Harald Havir\orcid{0000-0002-2944-986X}}
\affiliation{NanoLund and Solid State Physics, Lund University, Box 118, 22100 Lund, Sweden}
\author{Pierre Glidic\orcid{0000-0002-3099-4611}}
\affiliation{NanoLund and Solid State Physics, Lund University, Box 118, 22100 Lund, Sweden}
\affiliation{Université Lyon 1, CNRS, Institut Lumière Matière, UMR5306, F-69100, Villeurbanne, France}
\author{Samuel Andersson\orcid{0009-0004-3818-4420}}
\affiliation{NanoLund and Solid State Physics, Lund University, Box 118, 22100 Lund, Sweden}
\author{Sebastian Lehmann\orcid{0000-0002-4091-905X}}
\affiliation{NanoLund and Center for Analysis and Synthesis, Lund University, Box 124, 22100 Lund, Sweden}
\author{Kimberly A. Dick\orcid{0000-0003-4125-2039}}
\affiliation{NanoLund and Center for Analysis and Synthesis, Lund University, Box 124, 22100 Lund, Sweden}
\author{Claes Thelander\orcid{0000-0002-9794-1864}}
\affiliation{NanoLund and Solid State Physics, Lund University, Box 118, 22100 Lund, Sweden}
\author{Ville F. Maisi\orcid{0000-0002-2944-986X}}
\email{ville.maisi@fysik.lu.se}
\affiliation{NanoLund and Solid State Physics, Lund University, Box 118, 22100 Lund, Sweden}

\begin{abstract}
Charge detection of nanoscale quantum devices commonly relies on dissipative sensors, where signal absorption causes measurement backaction. Recently, nonlinear resonators showed a near-unity readout signal through an induced frequency shift. We show that this frequency shift protects the sensor, as it detunes the resonator from the measurement signal. The resonator excitation reduces while an enhanced readout signal is maintained. Our results, supported by numerical modeling, show that nonlinear resonators can combine strong readout contrast with reduced measurement-induced backaction.
\end{abstract}
 
\date{\today}
\maketitle


Resolving the charge state of a nanoscale quantum device is a central capability in both quantum technologies and fundamental studies of electron transport. Fast and sensitive charge detection enables, for example, the readout of semiconductor spin qubits, electron counting experiments, and investigations of single-electron dynamics~\cite{Schoelkopf1998,Pekola2013,Chatterjee2021,Burkard2023,Gustavsson2009}. To achieve high measurement bandwidth, charge sensors such as single-electron transistors, quantum point contacts, and quantum dots are commonly embedded in radio-frequency resonator circuits, where changes in sensor conductance are converted into changes in the reflected microwave signal~\cite{Reilly2007,Barthel2010,Vigneau2023,Harabula2017,deJong2021}.

The resonator enables impedance matching between the sensor-induced dissipation and the resonator-input coupling~\cite{Vigneau2023,Ares2016}. This yields the maximal signal with all the incident microwave signal transmitted and dissipated in the sensor. Increasing the power of the readout signal improves further the capability to resolve the sensor state efficiently. Dissipating all the signal in the sensor and increasing the signal, however, perturbs the sensor and the measured system through induced backaction, such as broadening of the transport features~\cite{Frey2011}, rectification~\cite{Cornia2019}, photon-assisted tunneling~\cite{Mavalankar2016}, detector-induced dephasing~\cite{Haldar2025}, and local heating~\cite{Keith2019}. These backaction effects limit the maximum usable signal and hence also set a limit to the maximum attainable speed.

Recently, near-unity charge-readout signals have been demonstrated without satisfying the conventional matching condition, as the dissipative response of the sensor instead produces a frequency shift of the nonlinear resonator similarly as for superconducting qubits~\cite{Siddiqi2004,Wallraff2005,Lupascu2006,Havir2026}. Consequently, the incident microwave power does not have to be fully absorbed. In this work, we demonstrate experimentally that the induced frequency shift protects the sensor from the readout signal. The shift detunes the resonator mode from the drive and hence lowers the voltage experienced by the sensor. We observe the arising protection as a decrease in the rectification current of the sensor, despite of the incident power increasing. This yields a protected readout scheme that reduces the backaction while the readout signal still enhances and attains a near-unity value at the same time.

We demonstrate the protected readout with a nonlinear SQUID-array resonator coupled directly to a quantum dot. We show that the frequency shift reduces the resonator voltage experienced by the quantum dot, thereby suppressing the drive-induced current. At the same time, the readout signal remains strongly enhanced. We also present numerical modeling that reproduces the effects observed in the experiments qualitatively and confirms the picture that the drive amplitude is indeed suppressed thanks to the resonator shift. These results provide a route towards fast and minimally invasive charge detection.

\begin{figure*}[t]
    \centering
    \includegraphics[width=\textwidth]{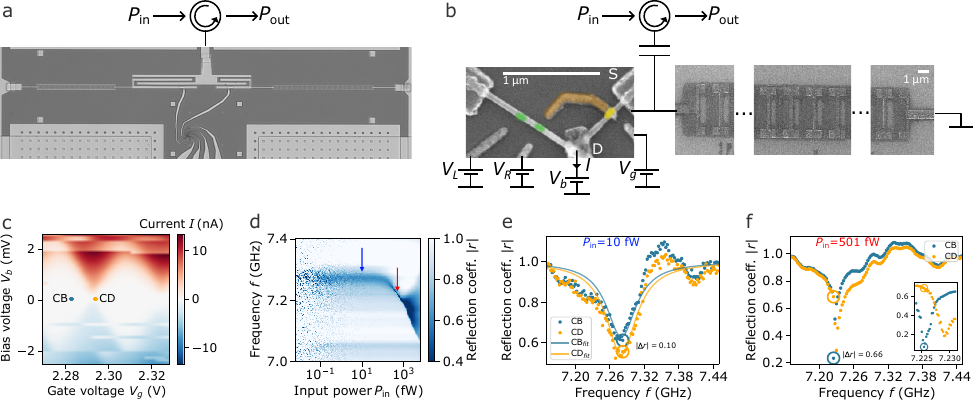}
    \caption{(a) Optical image of the device. (b) Scanning electron micrograph of the QDs. The single quantum dot (SQD), metallic coupler and double quantum dot (DQD) are indicated in yellow, orange and green, respectively, together with the source and drain leads. The SQD is coupled to a nonlinear resonator formed by an array of $N=61$ SQUIDs. 
    (c) Current $I$ through the SQD as a function of gate voltage $V_g$ and bias voltage $V_b$, showing a Coulomb degeneracy at $V_g = 2.295$~V. 
    (d) Reflection coefficient $|r|$ as a function of input power $P_{\mathrm{in}}$ and drive frequency $f$, illustrating the transition from linear to nonlinear resonator response. The blue/red arrow indicates the linear/nonlinear operation point at 10/501 fW. The SQD is in Coulomb blockade. (e)/(f) Reflection coefficient $|r|$ as a function of frequency $f$ for low/high input power, $P_{\mathrm{in}} = 10$~fW/$501$~fW, corresponding to the linear/nonlinear operation point. The SQD is tuned between Coulomb blockade (CB) and Coulomb degeneracy (CD). In the linear/nonlinear case, the difference between CB and CD leads to a change in the reflection coefficient $|\Delta r|=0.10/0.66$ marked by the open circles. The inset in (f) is a zoomed-in measurement of the main panel.
    }
    \label{fig:dev}
\end{figure*}

Our device, shown in Fig.~\ref{fig:dev}(a) and (b), has the same structure as previously employed in Ref.~\citealp{Havir2026}. It consists of a single quantum dot (SQD) sensor capacitively coupled to a double quantum dot (DQD) with a metallic coupler, indicated in yellow, green and orange, respectively. The electrochemical potentials of the dots are controlled via plunger gates $V_g$, $V_R$, and $V_L$. The DQD is not involved in the measurements presented here. Transport through the SQD is measured by applying a bias voltage $V_b$ and recording the current $I$, as shown in Fig.~\ref{fig:dev}(b). A representative Coulomb degeneracy (CD) is observed at gate voltage $V_g=2.295$~V with a conductance $G=0.15~\mu$S, as shown in Fig.~\ref{fig:dev}(c). We use this as the operating point for the measurements. All measurements are performed in a dilution refrigerator with an electronic base temperature of $T=50$~mK.

The SQD is coupled to a nonlinear microwave resonator formed by an array of $N=61$ SQUIDs. One end of the resonator is grounded, while the other is connected to the SQD, allowing the resonator voltage to directly modulate the source potential of the dot. Figure~\ref{fig:dev}(d) shows the reflection coefficient $|r|$ of the bare resonator as a function of input power $P_{\mathrm{in}}$ and drive frequency, illustrating the transition from linear to nonlinear response, consistent with Ref.~\citealp{Havir2026}. Figure~\ref{fig:dev}(e) shows further the resonator response $|r|$ in the linear regime with $P_{\mathrm{in}}=10$~fW (blue arrow in Fig.~\ref{fig:dev}(d)), for the SQD tuned either to Coulomb blockade (CB), $V_g=2.283$~V, or Coulomb degeneracy (CD), $V_g=2.295$~V, with a maximum difference in readout of $|\Delta r| = 0.1$. Figure~\ref{fig:dev}(f) correspondingly shows the resonator response in the nonlinear regime with $P_{\mathrm{in}}=501$~fW (red arrow in Fig.~\ref{fig:dev}(d)), for both SQD tuned to CB or CD. In this regime, the resonator response differs qualitatively from the linear case of Fig.~\ref{fig:dev}(e). In particular, the change in reflection amplitude increases to $|\Delta r| = 0.66$, compared to $0.10$ in the linear regime.

In the linear regime, the resonator response corresponds to the solid lines of Fig.~\ref{fig:dev}(e) with \cite{Havir2026}
\begin{equation}
    r = 1 - \frac{\kappa_c}{\kappa /2 - i(\omega - \omega_r)}.
    \label{eq:lorenz}
\end{equation}
Here $\omega_r$ denotes the resonance frequency of the resonator and $\kappa=\kappa_c+\kappa_i+\kappa_{\mathrm{QD}}$ the total linewidth of the mode, consisting of contributions from the coupling to the feedline $\kappa_c$, internal losses $\kappa_i$, and the dissipation induced by the quantum dot $\kappa_{\mathrm{QD}}$. Using this expression, we extract $\kappa_c/2\pi=57$~MHz, $\kappa_i/2\pi=13$~MHz at CB ($\kappa_{\mathrm{QD}}=0$), and an additional $\kappa_{\mathrm{QD}}/2\pi=4$~MHz at the CD by fitting Eq.~(\ref{eq:lorenz}) to the measured data. For the CD case, $\kappa_{\mathrm{QD}}$ is the only free parameter. As $\kappa_{\mathrm{QD}}\ll\kappa_c$, the SQD dissipation is too small to reach matching, resulting in a small change in the reflection amplitude.

The added dissipation of $\kappa_{\mathrm{QD}}$ of the SQD results primarily in a change in linewidth for the linear regime, leading to a small signal contrast. In contrast, in the nonlinear regime, the added dissipation leads to a frequency shift instead, producing a significantly larger change in the reflected signal. This enhancement arises from the nonlinear conversion of the dissipation induced by the SQD into a frequency shift of the resonator, rather than a simple broadening of the resonance. As a result, the readout signal becomes both stronger and more sensitive to the charge state of the system, as demonstrated in Ref.~\citealp{Havir2026}.

\begin{figure}
    \centering
    \includegraphics[width=0.45\textwidth]{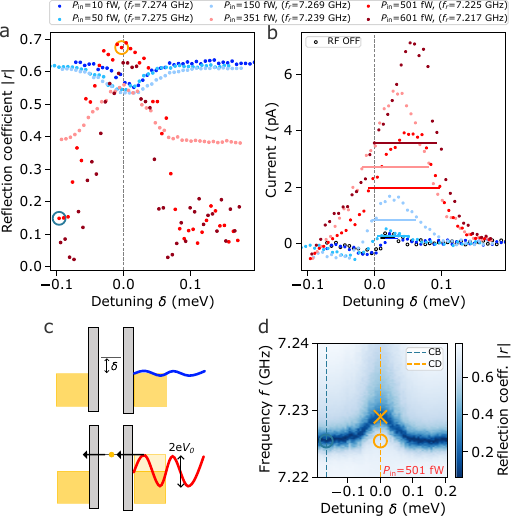}
    \caption{
    (a)/(b) Reflection coefficient $|r|$/current $I$ through the SQD as a function of detuning $\delta$ from the Coulomb degeneracy for increasing input powers $P_{\mathrm{in}}$.
    (c) Schematic illustration of the effect of the resonator voltage on the SQD. For small drive amplitudes (blue), the voltage oscillations are negligible and transport occurs only near Coulomb degeneracy. For large drive amplitudes (red), the oscillating voltage $V_0$ modulates the source potential over a finite range, enabling transport over a wider detuning window. (d) Reflection coefficient $|r|$ as a function of frequency $f$ and detuning $\delta$ from Coulomb degeneracy at $P_{\mathrm{in}} = 501$~fW. The blue circle indicates the bare resonator response in CB while the orange circle/cross indicates the protected/unprotected operating point. The dashed lines labeled CB and CD are the linecuts shown in Fig.~\ref{fig:dev}f.}
    \label{fig:broad}
\end{figure}

Next we consider the backaction effects. The backaction causes electrical current driven through the SQD to increase \cite{Cornia2019, Mceuen1996} and the linewidth of the resonance peak to broaden \cite{Frey2011} with increasing drive. Figures~\ref{fig:broad}(a),(b) show these effects for our device by plotting the reflection coefficient $|r|$ and the current $I$ through the SQD as a function of detuning $\delta$ around the Coulomb degeneracy for increasing input powers. For each input power, the drive frequency is chosen from the strongest resonator response in CB in Fig.~\ref{fig:dev}(d). In the absence of microwave drive, the current exhibits an oscillatory lineshape similar to a thermoelectric response. Such a lineshape arises from a temperature difference across the contacts~\cite{Svilans2016, Scheibner2007, Beenakker1992}, or from voltage fluctuations on the source and drain electrodes \cite{Cornia2019, Kouwenhoven1994}. With increasing power, the rectification process gets stronger, leading to increasing current as seen in Fig.~\ref{fig:broad}(b). Interestingly, only the response with positive polarity increases when the resonator becomes nonlinear at $P_{\mathrm{in}} > 100$ fW.

The broadening of the resonance peak and the increase of the maximum current continue up to the power $P_{\mathrm{in}} = 351$~fW (light red). Figure~\ref{fig:broad}(c) illustrates how the increasing voltage amplitude gives rise to the broadening: at low drive powers, the voltage oscillations in blue are small such that no transport takes place via a high lying energy level. In contrast, at higher drive powers in red, the oscillating resonator voltage raises the electrochemical potential of the source lead such that the high-lying SQD level enters the transport window during part of each cycle. As a result, the current peak broadens and the current increases, consistent with previously reported microwave-induced power broadening in quantum-dot transport measurements \cite{Cornia2019,Frey2011}. At the same time as the above backaction effects take place in the electrical current, the resonator response of Fig.~\ref{fig:broad}(a) also broadens.

In addition to the broadening, another effect takes place in the resonator response of Fig.~\ref{fig:broad}(a). At low power, the response is a dip and the decrease in $|r|$ originates from the increased SQD-induced losses, as seen in Fig.~\ref{fig:dev}(e). At higher drive powers, the resonator response reverses as the SQD is tuned through the Coulomb degeneracy. Notably, the response increases as it reverses. The change from a dip to a peak arises because the SQD-induced response shifts the resonance frequency in the nonlinear regime, rather than primarily broadening the resonance. As a result, the CD state becomes off resonant at the fixed drive frequency and approaches full reflection, explaining the increase in $|r|$. This frequency shift is shown directly in Fig.~\ref{fig:broad}d, where the resonance shifts increasingly away from the fixed drive frequency (blue circle) as the SQD approaches the Coulomb degeneracy, until it is fully detuned (orange circle).

\begin{figure}[t]
    \centering
    \includegraphics[width=0.45\textwidth]{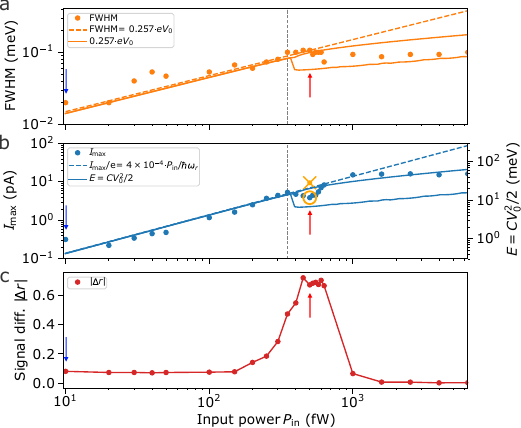}
    \caption{
    (a)/(b) The width of the current peak (FWHM)/maximum current $I_{\mathrm{max}}$ through the SQD as a function of input power $P_{\mathrm{in}}$ (orange/blue dots), together with the voltage amplitude $V_0$-based curves obtained from a numerical model (solid lines). The dashed lines are the low-power linear trends. The onset of the nonlinearities are indicated by the vertical dashed grey line. A pronounced dip in the maximum current is observed around $P_{\mathrm{in}} = 501$~fW, both in the measured data and in the corresponding low-amplitude branch of the numerical model.The orange circle/cross marks the current $I_{\mathrm{max}}$ measured at the protected/unprotected operating point. (c) The maximum signal difference $|\Delta r|$ between the two states CB and CD for increasing input drive $P_{\mathrm{in}}$. The highest difference is obtained around the protected operating point, indicated by the red arrow.}
    \label{fig:v}
\end{figure}

Figure~\ref{fig:broad}a shows that the light red trace marks the onset of the response reversal, while for the red trace the shifted response is fully developed. As the power is increased from the light red to red trace in Fig.~\ref{fig:broad}b, the current drops. This is a signature of suppressed backaction. Since the current suppression coincides with the nonlinear frequency shift, we attribute the reduced backaction to weaker driving of the resonator. We identify the operating point marked by the orange circle in Fig.~\ref{fig:broad}d as the protected operating point. Importantly, the reduction in resonator voltage does not come at the expense of the readout signal. The protected operating point combines reduced measurement-induced backaction with a strongly enhanced readout signal. Figures~\ref{fig:v}b,c further show $I_{\mathrm{max}}$ and $|\Delta r|$ as functions of $P_{\mathrm{in}}$. The readout signal is maximized at the same input power for which $I_{\mathrm{max}}$ reaches its minimum in the dip.

The low-power regime of Fig.~\ref{fig:v} illustrates the role of the voltage amplitude in the response. The full width at half maximum (FWHM, orange points) of the current peak increases in proportion to approximately half of the resonator voltage amplitude $V_0/2$, within an uncertainty of less than a factor of two. This agrees with the picture in Fig.~\ref{fig:broad}c, where the resonator voltage sets the energy range accessible to the electron during the tunneling cycle, with approximately half of the voltage dropping across each tunnel barrier \cite{Cornia2019, Haldar2023}. The difference to this likely arises from calibration uncertainty of the input line attenuation power and additional capacitances to the QD, for example from the coupler, that lower the coupling of the resonator voltage. On the other hand, the maximum current $I_{\mathrm{max}}$ of Fig.~\ref{fig:v}b initially increases linearly with input power, reflecting the growing photon flux into the resonator, consistent with photon-assisted tunneling~\cite{Ingold1992,Souquet2014}. Expressed as the probability for an incident photon to result in a tunneled electron, this gives
$\eta=\frac{I/e}{P_{\mathrm{in}}/\hbar\omega_r}\approx4\times10^{-4}$.
The low efficiency can be attributed to the lack of directionality for electron transport through the SQD, in comparison to the directivity built in to a DQD system yielding 6-68\% \cite{Khan2021, Haldar2025_2, Oppliger2026}.

To further consolidate the voltage-amplitude-based backaction, we consider the circuit model of Ref.~\citealp{Havir2026}. The solid lines in Fig.~\ref{fig:v}b show the energy $E = C V_0^2/2$ stored in the resonator, where $C$ is the total capacitance of the resonator. Here we extended the model to determine the voltage amplitude $V_0$ in the resonator by using the Josephson relation
\begin{equation}
    V_0 = N\frac{\Phi_0}{2\pi}\frac{d\phi}{dt},
\end{equation}
where $\phi$ is the superconducting phase across each junction and $\Phi_0 = h/2e$ is the magnetic flux quantum. All model parameters are obtained from the low-power measurements without introducing additional adjustable parameters. In the bifurcation regime, the numerical model yields low- and high-amplitude solutions, shown by the two solid lines. The calculated energy of the lower branch exhibits a dip at the same operating point as the measured maximum current, whereas the higher branch follows the overall increasing trend. Fluctuations in the resonator amplitude $V_0$ induce switching between the two states, resulting in an averaged response of the two branches~\cite{Siddiqi2004}. Such fluctuations are not included in the model, which therefore yields the two branches separately. The depth of the measured dip is however comparable to the separation between the two calculated branches. This indicates that near the bottom of the dip the resonator predominantly occupies the lower-amplitude branch, corresponding to the protected state, whereas outside the dip it predominantly occupies the higher-amplitude, unprotected branch. The agreement between the measured and calculated responses therefore supports the interpretation that the current suppression originates from a reduction of the resonator voltage.

We can further show the disappearance of the protection by driving the resonator in resonance. We drive at the resonance frequency of the protected response, marked by the orange cross in Fig.~\ref{fig:broad}d. Under this condition, the frequency detuning responsible for the protection is removed. As a result, the measured maximum current through the SQD is larger and aligns with the linear trend with input power, as shown by the orange cross in Fig.~\ref{fig:v}b. This indicates that resonant driving efficiently excites the resonator, maintaining a high voltage amplitude and thereby restoring the backaction on the SQD.

In summary, we have shown a detector operation mode where a charge sensor is protected from the readout signal. The protection arises as dissipation induces a frequency shift to a nonlinear resonator, the same effect that produces a strongly enhanced readout signal. The sensor is protected from the measurement signal as the frequency shift detunes the resonator off from the fixed measurement drive frequency. This off-resonant condition results in lower resonator excitation and thereby suppresses the drive-induced backaction current. We further show that the suppressed backaction takes place at the same time as the readout signal is enhanced to its maximum value. This combination of strong readout contrast and reduced detector-induced disturbance makes nonlinear resonators attractive for fast charge readout of nanoscale quantum devices. Furthermore, the demonstrated protection here applies more generally for any dissipative sensors.

\bibliography{master}
\bibliographystyle{apsrev4-2}

\end{document}